\documentclass[superscriptaddress,twocolumn,amsmath,amssymb,aps,pre,floatfix]{revtex4-2}
\usepackage{comment,color}
\usepackage[english]{babel}
\usepackage[utf8]{inputenc}
\usepackage[T1]{fontenc}
\usepackage{booktabs}
\usepackage{xr,comment}
\usepackage{float}
\usepackage{natbib}
\usepackage{units}
\usepackage{graphicx}
\usepackage{subcaption}
\usepackage{animate}
\usepackage{enumitem}
\usepackage{dcolumn}
\usepackage{bm}
\usepackage{hyperref}
\hypersetup{colorlinks=true, linktoc=all, linkcolor=blue, linktocpage, citecolor=blue}

\renewcommand{\aa}{\mathrm{a}}
\newcommand{\bb}{\mathrm{b}}
\newcommand{\cc}{\mathrm{c}}
\newcommand{\abs}{\mathrm{abs}}
\newcommand{\esc}{\mathrm{esc}}
\begin{document}

\title{
Cross-feeding percolation phase transitions of inter-cellular metabolic networks}
\author{Luís C. F. Latoski}
\email{luis.latoski@ufrgs.br}
\affiliation{Instituto de Física, Universidade Federal do Rio Grande do Sul, CEP 91501-970, Porto Alegre - RS, Brazil}
\author{Andrea De Martino}
\email{andrea.demartino@polito.it}
\affiliation{Politecnico di Torino, Corso Duca degli Abruzzi 24, 10129 Torino, Italy}
\affiliation{Italian Institute for Genomic Medicine, SP142 Km 3,95, 10060 Candiolo, Italy}
\author{Daniele De Martino}
\email{daniele.demartino@ehu.eus}
\affiliation{Biofisika Institute (CSIC, UPV/EHU), Barrio Sarriena s/n, 48940 Leioa, Bizkaia, Spain}
\affiliation{Ikerbasque Foundation, Bilbao, Spain}

\begin{abstract}
Intercellular exchange networks are essential for the adaptive capabilities of populations of cells. While diffusional exchanges have traditionally been difficult to map, recent advances in nanotechnology enable precise probing of exchange fluxes with the medium at single-cell resolution. Here we introduce a tiling-based method to reconstruct the dynamic unfolding of exchange networks from flux data, subsequently applying it to an experimental mammalian co-culture system where lactate exchanges affect the acidification of the environment. We observe that the network, which initially exhibits a dense matrix of exchanges, progressively breaks up into small disconnected clusters of cells. To explain this behaviour, we develop a two-parameter Maximum-Entropy multicellular metabolic model that incorporates diffusion-driven exchanges through a set of global constraints that couple cellular behaviors. The model predicts a transition from a densely interconnected network to a sparse, motif-dominated state as glucose and oxygen consumption levels shift. We characterize such a crossover both numerically, revealing a power-law decay in the cluster-size distribution at the critical transition, and analytically, by computing the critical line through a mean-field approximation based on percolation theory. By comparing empirical data with theoretical predictions, we find that   populations evolve towards the sparse phase by remaining near the crossover point between these two regimes. These findings offer new insights into the collective organization driving the adaptive dynamics of cell populations.
\end{abstract}

\maketitle

\section*{Introduction} 

The molecular revolution of the last century led to multiple breakthroughs in biology, culminating with the advent of whole-genome sequencing and its promises for personalized medicine. However, it also encountered a challenge: is it possible to reconstruct complex physiological behavior from highly heterogeneous molecular interactions \cite{ball2023life}? In response to this challenge, the ``systemic'' approach of systems biology shifted the focus from the nodes (molecules) to their interactions \cite{palsson2015systems,klipp2016systems,alon2019introduction}. While shedding light on many previously unseen aspects of biological systems, these methods often result in {\it ad hoc} models overloaded with molecular details, capable of reproducing empirical data but with limited generalization potential. By contrast, statistical mechanics has successfully explained macroscopic behavior by connecting molecular interactions and large-scale phenomena through highly generalizable frames in somewhat simpler physical systems, such as gases, magnets or metals \cite{reichl2016modern}, but also on more complex ones like glasses \cite{leuzzi2007thermodynamics} or the brain \cite{amit1989modeling}, finding broad applicability in biophysics as well \cite{bialek2012biophysics}. At least in principle, the conceptual toolbox of statistical mechanics might be suited to yield new outlooks on systemic biological problems.

One of the major additional difficulties posed by living systems lies in the fact that they operate across a hierarchy of scales, from single molecules to ecosystems, with the cellular level arguably playing the central role. There is therefore a pressing need for research that addresses emergent phenomena across multi-scale interaction structures, particularly at the boundary between the molecular and cellular levels. In this article we address one such structure, arising in the context of overflow metabolism. 

The terms `overflow metabolism' refer to the ubiquitous and still not fully understood phenomenon whereby cells, even in the presence of sufficient oxygen, prioritize less efficient metabolic pathways (like fermentation in yeast or aerobic glycolysis in cancer cells) over more efficient ones, leading to the excretion of metabolic byproducts such as lactate \cite{vazquez2017overflow}. This behavior is typically observed when cells are rapidly proliferating and is usually ascribed to nutrient uptake that exceeds the capacity of oxidative metabolism, in turn allowing for faster energy production at the cost of efficiency. The latter scenario has been studied in detail in bacteria \cite{basan2015overflow,mori2016constrained,mori2019yield,de2020common}. Despite numerous studies, its validity for mammalian cells, where inter-cellular interactions are likely to play an important role, is still unclear \cite{vander2009understanding,vazquez2010catabolic,schuster2015mathematical,deberardinis2020we} 

At odds with the cell-autonomous scenario, lactate accumulation in the medium (also known as Warburg effect, the key sign of overflow) could result from (a breakdown of) the collective organization of inter-cellular exchanges \cite{narayanankutty2024emergent}. Cells adapting to an environment indeed exchange chemical compounds for reasons ranging from signaling to cross-feeding. Exchanged compounds are usually byproducts of the cells' metabolic activity that are excreted in the medium by one cell and are in-taken by another cell upon being sensed by specialized surface receptors. Uptake in turn affects the importer's metabolic activity, leading to a modulation of its secretions, i.e. to a new batch of signals that propagate across the population and trigger uptakes by other cells. As exchanges and interactions build up, they effectively establish a degree of decentralized cross-cellular coordination that facilitates the overall adaptation of the population to the medium and enable to engineer a viable (shared) environment. It is now known that overflow induces cell-to-cell interactions driven by exchanges of lactate \cite{onesto2022probing}. By balancing lactate excretion and import rates, populations can in principle control environmental lactate levels even in presence of cells with sustained excretion rates. By contrast, a failure in this process can lead to the Warburg phenomenon. The structure and dynamics of exchange networks therefore potentially encode a much more resolved picture of how a population of cells adjusts to and shapes its surrounding environment than the one provided, for instance, by macroscopic quantities such as the lag (adaptation) times in traditional growth curves.

Up to recent times not much was known about these networks, although some of their properties were easy to guess. From a physics viewpoint, in fact, when cells have no mechanical contact with each other and no advective transport is present in the medium (as in the case we consider here), the emergent interaction network is sustained by the diffusive motion of signaling molecules. This implies that interaction networks are bound to be intrinsically stochastic, potentially lossy (due to signals being `perishable', in that the interaction-carrying molecule, once injected by a cell into the environment, could leave the system without being absorbed by another cell), dynamic (i.e. time-dependent) and highly heterogeneous (as exchanges between distant cells are far less likely than exchanges between nearby cells). 

Recent development in nanoscale technology applied to cellular systems  {\it in vitro}, however, have finally shed some light on these networks. In particular, techniques ranging from nanoscale mass-spectrometry \cite{musat2016tracking} to environment sensing through fluorescent nanofibers \cite{onesto2022probing} now allow for direct or indirect quantification of single-cell exchange fluxes with the medium (i.e. the rates at which cells excrete or import certain compounds) even at high spatial- and temporal resolution. These fluxes are the key quantities from which exchange networks can be reconstructed. 

Here we study how overflow is modulated by the underlying inter-cellular exchange network, showing theoretically that the latter can undergo a percolation transition from a dense to sparse phase that, in turn, could rescue the system from overflowing. We will show how empirical data suggest that populations adapting to an environment evolve in time towards the sparse phase by staying close to the transition line, through the gradual dilution of an initially dense exchange matrix.

The article is organized as follows. The first question we tackle is how to reconstruct the exchange network from single-cell flux data. Using the method developed to address this problem, we find a dense-to-sparse crossover in time as cells in an experimental population adapt to the medium. Next, we will define a two-parameter Maximum-Entropy constraint-based metabolic network model that reproduces the reconstructed exchange web quantitatively from the activity of single cells as an emergent feature. Finally, we will show that the dense-to-sparse crossover found in empirical data corresponds to a percolation phase transition in the model, with a well defined critical behavior. Biological implications are discussed in the final part of the article.

\section*{Results} 

\subsection*{Network reconstruction problem}

The experiments we consider are based on recently-developed techniques to probe cellular microenvironments \cite{grasso2023fluorescent}. More specifically, we focus on space- and time-resolved maps of pH levels obtained through the deployment of arrays of ratiometric fluorescent microsensors across a cell culture \cite{rizzo2022ph}. 
In short, pH values measured over time at a large number of points across the culture can be used to infer --for each cell in the population-- the net import or export flux of protons upon assuming that local pH levels (a) equilibrate faster than a few minutes (the temportal resolution of the experiments), 
and (b) are only affected by inter-cellular exchanges of acids (e.g. lactate) and by the buffering activity of the medium. The specific case we analyze concerns the early adaptation (lag phase) of a mixed population of human pancreatic cancer cells and cancer-associated fibroblasts sharing the same carbon- and oxygen-rich medium, in which single-cell proton fluxes obtained via statistical inference could be linked directly to lactate uptakes and outtakes \cite{onesto2022probing}. For our modeling work, we reduced the setup to the bare essentials (see below). 
The reader is however referred to \cite{onesto2022probing} for  experimental and computational details of the background empirical picture on which we rely.

Our modeling frame is as follows. In an $L\times L$ square (corresponding to the observable frame, where $L\sim 0.5$ mm and where we neglect the $z$-direction that is present in experiments since cells adhere to the substrate), $N\sim 150$ cells of radius $R\ll L$ ($R\sim10\,\mu$m) are placed at fixed positions $\mathbf{r}_i$ (in agreement with the highly reduced mobility seen empirically \cite{onesto2022probing}). For each cell, a flux variable $u_i(t)$ (in units of millimols per gram of dry weight per hour: mmol/gh) is given, corresponding to the net import ($u_i>0$) or export ($u_i<0$) of a certain compound over a (sufficiently large) time window ending at time $t$. This compound travels through the square by ordinary diffusion. The task is to reconstruct the inter-cellular exchange network at all time steps, i.e. find time-dependent exchange fluxes $u_{i\to j}>0$ such that
\begin{equation}
\sum_{j:u_j>0} u_{i\to j}\leq -u_i~~~
\end{equation}
if $u_i<0$ (i.e. if cell $i$ is an exporter, where the difference between the two sides of the inequality stands for the net flux accumulating in the medium and/or leaving the frame), and
\begin{equation}
\sum_{j:u_j<0} u_{j\to i}=u_i
\end{equation}
if $u_i>0$ (cell $i$ is an importer). Equivalently, we would like to compute the probability $P(j|i)$ that a molecule absorbed by a cell at position $\mathbf{r}_j$ was excreted by a cell at position $\mathbf{r}_i$, as inter-cellular exchange fluxes are given by 
\begin{equation}\label{vfvgsdwef}
    u_{i\to j}=u_j P(j|i)~~.
\end{equation}

In principle, because of the assumed separation of time scales, the matrix $P(j|i)$ can be found by solving a system of interconnected exit problems, i.e. Laplace equations with complex boundary conditions defined by cells and by the system's boundary \cite{medkova2018laplace}. Since diffusion is the only transport mechanism involved, however, it is reasonable to expect that $P(j|i)\propto 1/d_{ij}$, where $d_{ij}=|\mathbf{r}_i-\mathbf{r}_j|$ is the distance between the cell at point $i$ and that at point $j$. A simple pairwise assumption would indeed suggest that 
\begin{equation}\label{mfu}
    u_{i\to j} \simeq -\frac{1}{Z_i}\,\frac{u_i u_j}{d_{ij}}~~,
\end{equation}
where $u_i<0$ ($i$ is exporting), $u_j>0$ ($j$ is importing), and 
\begin{equation}
    Z_i =  \sum_{k:u_k>0} \frac{u_k}{d_{ik}}
\end{equation}
is a normalization factor. The exact solution, on the other hand, can be obtained through a Monte Carlo method that simulates the two-dimensional diffusion process (``walk on spheres'', \cite{muller1956some}, see Appendix). For our extensive numerical analysis, we however devised a computationally less demanding alternative that, as we shall see, yields quantitatively comparable results.

\subsection*{Stochastic Voronoi approach and empirical networks}

\begin{figure*}
    \centering
    \includegraphics[width=\textwidth]{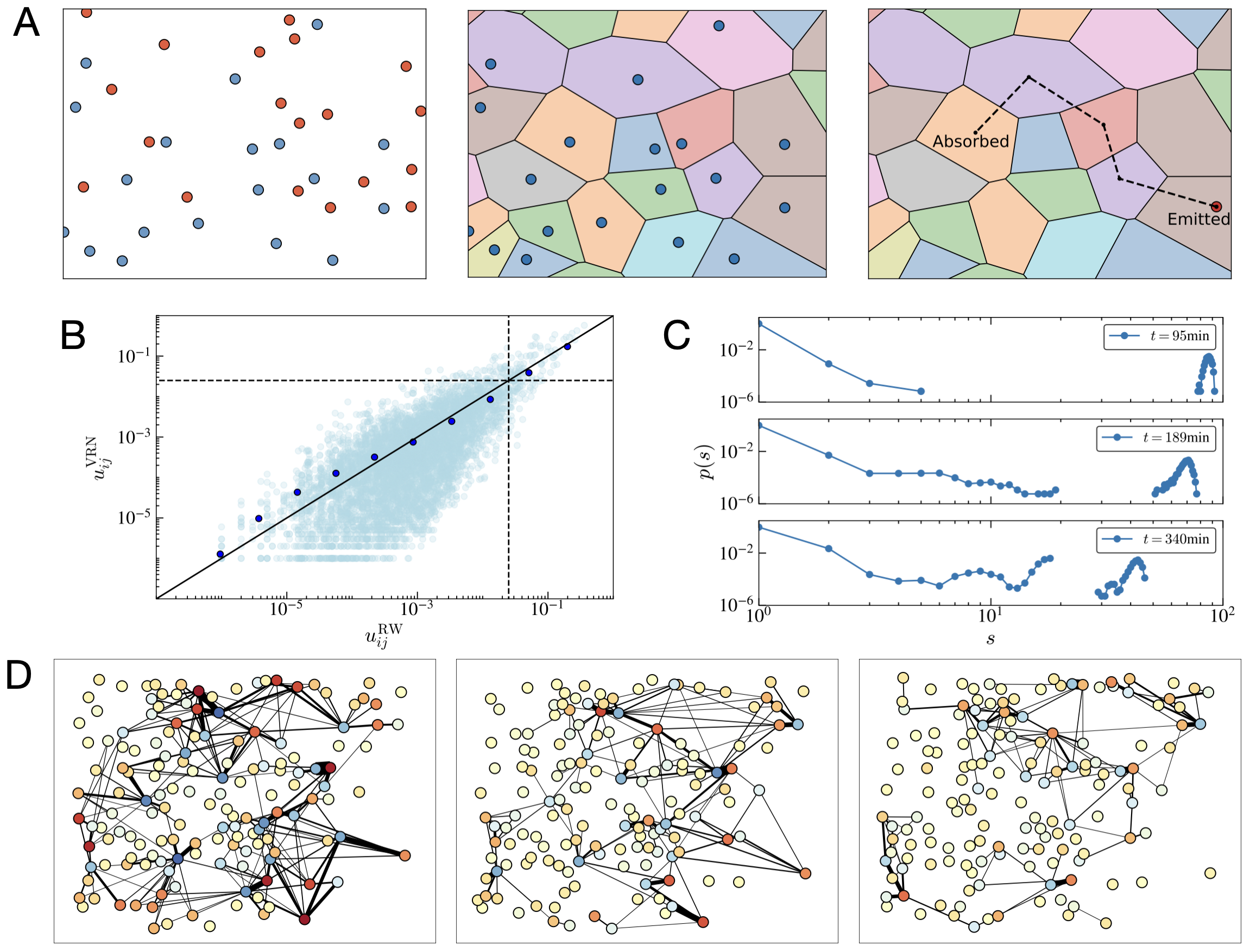}
\caption{{\bf Reconstruction of the experimental cross-feeding network during the Warburg effect.} (A) Given the positions of cells sharing a certain environment (red being emitters, blue being absorbers), we construct a Voronoi tessellation using absorbers as seeds. The algorithm we employ allows to calculate the probability that a particle secreted by an emitter cell gets absorbed in any given tile. (B) Comparison between results for exchange fluxes obtained via the Voronoi-based method and the (exact) Monte Carlo method described in the Appendix, for the first frame from the dataset given in \cite{onesto2022probing}. The dashed lines mark the threshold $\theta=0.025$ mmol/gh, such that cells with larger exchange fluxes are defined to be topologically connected. (C) Distribution of cluster sizes $(s)$ obtained from experiments from \cite{onesto2022probing} with $N\sim 150$ at times $t=95$, $t=189$ and $340\,\unit{min}$ (top to bottom). (D) Reconstructed networks based on the mean single-cell lactate exchange fluxes corresponding to the time stamps shown in panel (C). Note how the largest cluster of exchanges that initially dominates the population gradually dissipates as the population partitions into smaller exchange clusters.}
    \label{fig1}
\end{figure*}

We begin by partitioning the $L\times L$ square into Voronoi tiles starting from the positions of the absorbing cells \cite{aurenhammer2000voronoi} (Fig. \ref{fig1}A). Now focus on the tile of an absorbing cell ($u_i>0$), say it is formed by $n$ segments of length $\ell_k$ ($k=1,\dots, n$), and adopt the general line of reasoning of \cite{berg1993random}. If a molecule performs a random walk inside this tile, the probability $p_\bb$ that the walk hits the outer boundary of the tile is asymptotically given by the ratio between the overall length of the outer boundary  ($\sum_k \ell_k$) and the sum of the lengths of the outer ($\sum_k \ell_k$) and inner ($2\pi R$, with $R$ the cell's radius) boundaries: 
\begin{equation}
p_\bb=\frac{\sum_k \ell_k}{2\pi R+\sum_k \ell_k}~~.
\end{equation}
On the other hand, the probability $p_\cc$ that the molecule hits the seed of the Voronoi tile (i.e. the cell) is just
\begin{equation}
p_\cc=1-p_\bb=\frac{2\pi R}{2\pi R+\sum_k \ell_k}~~.
\end{equation}
If, once it hits the cell, the molecule is absorbed with probability $p_\aa$, then the total probability $p_\abs$ that it is absorbed within that tile is obtained by summing the probabilities of all walks that end up with an absorption event. Noting that the probability that a walk hits the cell $n$ times without being absorbed is $p_\cc^n(1-p_\aa)^n$, we have
\begin{equation}
p_\abs=p_\cc \,p_\aa\,\sum_{n\geq 0}p_\cc^n(1-p_\aa)^n=\frac{p_\cc \,p_\aa}{1-p_\cc(1-p_\aa)}~~,
\end{equation}
where the term $p_\cc p_\aa$ accounts for the probability that an absorption event follows $n$ non-absorption ones. As a consequence, the molecule escapes the tile by crossing one of its boundaries with probability $p_\esc=1-p_\abs$, and escape occurs through the $k$-th boundary segment with probability 
\begin{equation}
p(k|\esc) = \frac{\ell_k}{\sum_k \ell_k}~~.
\end{equation}

The complete random walk of a particle across the tiled square can therefore be seen as a Markov chain with $N+1$ states ($N$ states for `particle is in $r$-th tile with $r=1,\ldots,N$ plus one state, labeled `0', corresponding to `absorbed') and transition probabilities given by 
\begin{gather}
W(r\to s) = (1-p_{\abs,r}) \frac{\ell_{rs}}{L_r}~~,\\
W(r\to 0) = p_{\abs,r} ~~,
\end{gather}
where $r,s\in\{1,\ldots,N\}$, $p_{\abs,r}$ denotes the probability that the particle is absorbed in tile $r$,  $\ell_{rs}$ is the length of the segment shared by tiles $r$ and $s$ (equal to zero if $r$ and $s$ are not neighbours) and $L_r$ is the length of the outer boundary of tile $r$ ($L_r=\sum_s\ell_{rs}$). This implies that $P(j|i)$, the probability that a particle emitted by cell $i$ is absorbed by cell $j$, can be approximated numerically for all $j$ and $i$ by iterating the Markov chain just described a sufficiently large number of times using emitting cells as initial conditions for the random walks. Specifically,  
\begin{equation}
P(j|i)\simeq\frac{M(j|i)}{\sum_k M(j|k)}~~,
\end{equation}
where $M(j|i)$ is the number of molecules emitted at $i$ that are absorbed at $j$. 

The main advantage of this approach lies in its low computational cost compared e.g. to exact Monte Carlo methods: once the Voronoi tessellation is found, relevant quantities can be calculated straightforwardly. Fig. \ref{fig1}B shows that results obtained through the Voronoi method are in excellent agreement with those obtained from Monte Carlo when applied to a representative case (the first frame from the dataset in \cite{onesto2022probing}). Note how calculated exchange fluxes for empirical populations vary over roughly six orders of magnitude. Such a strong heterogeneity is a robust feature of empirical data. To focus the network structure on significant exchanges, however, we apply a threshold criterion on the value of $u_{i\to j}$: a connection between $i$ and $j$ will be assumed to be present only if $u_{i\to j}$ exceeds the average lactate flux at the overflow transition (see next section), which is found to be equal to $\theta = 0.025$ mmol/gh.   

The topologies of the resulting networks can be analyzed by different quantitative measures. The simplest possibility is to study the distribution of cluster sizes, whereby two cells are considered to be in the same cluster when they are either joined by a significant (above-threshold) exchange or by a chain of significant exchanges, independently of their direction. A few representative distributions are shown in Fig. \ref{fig1}C. One sees that the exchange network derived from the experimental dataset is characterized by the rapid formation, and subsequent slow decay, of a large component of connected exchanges involving, at its maximum, up to 70\% of cells (Fig. \ref{fig1}C). Such a behaviour is observed robustly in all realizations of the reconstructed network and $40\%$ of the single links can be reconstructed accurately (see the appendix). Snapshots displaying this scenario at three different times are shown in Fig. \ref{fig1}D. This dynamics can be compared to that of lactate spillover discussed in \cite{onesto2022probing}. Because lactate spillover decreases over time in experiments, densely-networked states are characterized by larger lactate spillover in the medium compared to more fragmented states, in which exchanges are well balanced and spillover is reduced.  


\begin{figure*}[ht!!!!!]
    \centering
    \includegraphics[width=\textwidth]{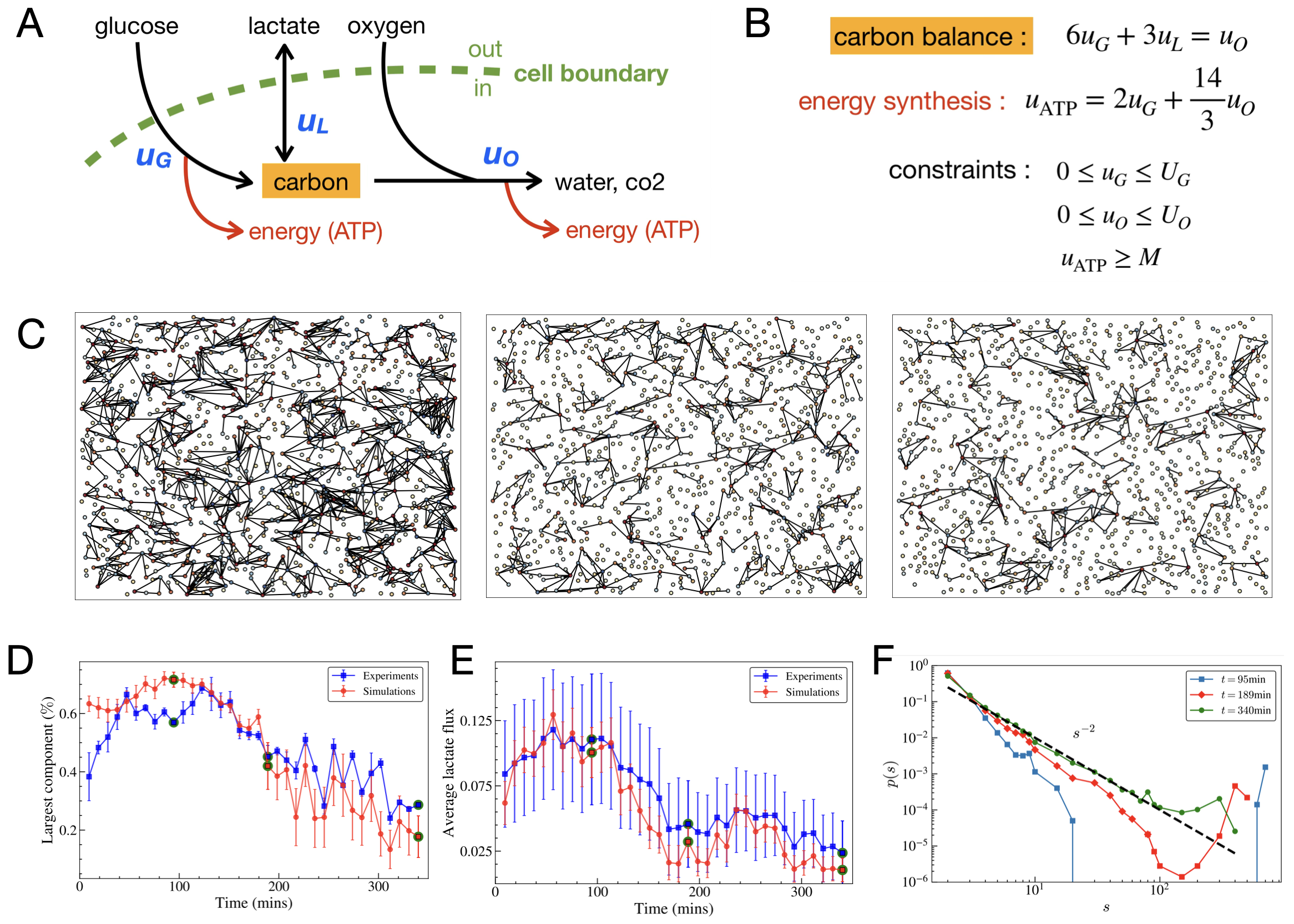}
\caption{ {\bf Metabolic network model recapitulates experimental data.} (A) Intracellular metabolic network model and (B) corresponding constraints enforcing carbon mass balance as well as bounds on import fluxes and on energy (ATP) synthesis for each cell. In the absence of competition for glucose and oxygen, the state of the $N$-cell system is determined by single-cell constraints plus the global constraint (\ref{eq:diff}) induced by lactate exchanges. (C) Representative exchange networks reconstructed from simulations of a population of cells (each carrying a metabolic network whose state on average is inferred from experiments) globally coupled by diffusion. Open boundary conditions are assumed. Results are shown for times $t=95$ min ($\beta_O = -0.7$, $\beta_G=-1.45$), $t=189$ min ($\beta_O = -1.4$, $\beta_G=-7.85$), and $t=340\,\unit{min}$ ($\beta_O = -1.9$, $\beta_G=-10.35$), same as Fig. \ref{fig1}. (D) Average fraction of nodes in the largest connected component found in experiments (blue) and computations (red) (averaged over $1000$ samples for each time frame). (E) Average lactate flux (Eq 13) found in experiments (blue) and model simulations (red) (averaged over $1000$ samples for each time frame). (F) Distribution of cluster sizes from model simulations for the same representative times in the experiments. The black dashed line displays the power-law behaviour $p(s)\sim s^2$. This is to be compared with the results from experiments, shown in Fig. \ref{fig1}C.}
    \label{fig2}
\end{figure*}

\subsection*{Constraint-based metabolic model}

In view of the fact that the activities of a large number of cells contribute to the formation and dynamics of the exchange network, the above results raise two broad theoretical questions. First, can the dense-to-sparse crossover in exchanges we observe in experiments be explained in terms of a transition in the collective behavior of the system? This would directly indicate that, despite being spatially separated, cells do not make fully autonomous metabolic choices but, rather, use diffusing signals to coordinate their behaviour, likely to achieve a better control of their microenvironments. Secondly, if so, how would such a coordination be connected with the metabolic activity of individual cells? 
To get some insight it is necessary to view the empirical exchange network as an emergent feature of the activities of a large number of metabolic networks (the individual cells) that respond to local cues like nutrient and pH levels. To implement such a `network of networks' approach, we studied the exchange networks generated by a minimal  model of a spatially-organized population of $N$ interacting metabolic networks, first introduced in  \cite{narayanankutty2024emergent}. 

Each cell $i$ ($i=1,\ldots,N$) is assigned a position $\mathbf{r}_i$ in a 2-dimensional square and its metabolism is assumed to be described by three variables, namely the glucose import flux $0\leq u_{G,i}\leq U_G$, the oxygen import flux $0\leq u_{O,i}\leq U_O$ and the flux of lactate exchange with the medium $u_{L,i}$, whereby $u_{L,i}>0$ if cell $i$ imports lactate and $u_{L,i}<0$ if cell $i$ exports lactate (Fig. \ref{fig2}A). 
The constants $U_G$ and $U_O$ stand for the finite capacity of cells to accommodate for glucose transporters and 
mitochondria, respectively.
At steady state, the three fluxes are related by a condition expressing carbon mass-balance, namely 
\begin{equation}
\label{eq:mass_balance}
u_{G,i}  +\frac{u_{L,i}}{2} = \frac{u_{O,i}}{6}~~~\forall i~~,
\end{equation}
which leaves two independent fluxes per cell. Moreover, the rate of energy (ATP) production by each cell is given by
\begin{equation}\label{eq:fatp}
u_{\mathrm{ATP},i} = 2u_{G,i}+\frac{14}{3}u_{O,i} ~~.
\end{equation}
To sustain survival, it is required that $u_{\mathrm{ATP},i}\geq M$ for all $i$, with $M$ a value representing the minimum necessary (maintenance) rate of energy production. Reaction fluxes of feasible single-cell metabolic states therefore satisfy the constraints shown in Fig. \ref{fig2}B. This model of a cellular metabolic network is a widely used minimal version of a constraint-based metabolic network \cite{majewski1990simple,vazquez2010catabolic,palsson2015systems,fernandez2017microenvironmental}, known to capture stylized features of more realistic and biochemically detailed models \cite{capuani2015quantitative}. 

When $N$ such networks interact in a shared medium without competing for external nutrients (to comply with the nutrient abundance employed in experiments), an effective coupling between cells can be induced by lactate exchanges. In absence of external sources of lactate, any spillover in the medium must indeed be endogenous, i.e. due solely to the activity of cells. This translates into a set of system-wide conditions on the lactate fluxes of different cells. Specifically, one finds that \cite{narayanankutty2024emergent}
\begin{equation}
\label{eq:diff}
\sum_{j=1}A_{ij}u_{L,j}\leq 0~~~\forall i~~,
\end{equation}
where 
\begin{equation}
A_{ij}=\delta_{ij}+(1-\delta_{ij})\frac{R}{|\mathbf{r}_i-\mathbf{r}_j|}
\end{equation}
are coefficients that account for the fact that lactate exchanges are diffusion-limited. In these conditions, a feasible state of the $N$-cell system corresponds to a pair of fluxes $(u_{G,i},u_{L,i})$ per cell that satisfies the $N$ constraints given above for single cells (Fig. \ref{fig2}B), plus the global constraint (\ref{eq:diff}). 

To explore the space of feasible states, we weigh $N$-cell configurations according to the Boltzmann factor
\begin{equation}\label{eq:bol}
e^{\beta_G\sum_i u_{G,i}+\beta_O\sum_i u_{O,i}}~~,
\end{equation}
where the auxiliary `inverse temperatures' $\beta_G$ and $\beta_O$ modulate the average flux state of the population. (We choose to represent states via the glucose and oxygen fluxes of each cell, which suffice to determine the missing lactate and energy fluxes). The choice (\ref{eq:bol}) is tightly related to the Maximum Entropy models of metabolism described in \cite{de2016growth,de2018statistical}, which represent the least biased models compatible with experimental data on average fluxes. Here, however, it serves the additional purpose of allowing for a compact representation of the metabolic state of the whole $N$-cell population, via the parameters $\beta_G$ and $\beta_O$. Values for the parameters $U_G$, $U_O$ and $M$  are taken from the literature (see Appendix). The Lagrange multipliers $(\beta_O,\beta_G)$ have instead been fitted to experimental data via maximum likelihood through simulations of the system on a grid, as explained in \cite{narayanankutty2024emergent}. 

Snapshots of the emergent exchange network found for a system with $N=10^3$ cells are displayed in Fig. \ref{fig2}C. The time evolution found in empirical data is qualitatively reproduced, as the dense exchange network that initially spans across the population gets gradually diluted into smaller disconnected clusters. Such a trend is also clear from the behaviour of the relative size (fraction of cells involved) of the largest exchange cluster (Fig. \ref{fig2}D) and from the population-averaged lactate flux (Fig. \ref{fig2}E), both of which quantitatively mimic the time trends retrieved from experimental data. Notice that, while the theoretical mean lactate flux ultimately yields a consistency check for the inference protocol, the relative size of the largest cluster is a genuine prediction of the model. The behaviour of the probability distribution of the sizes of clusters $p(s)$ is especially noteworthy (Fig. \ref{fig2}F). At earlier times, $p(s)$ is dominated by a single `giant' cluster. As time progresses, a large cluster persists (albeit reduced in size) along with a significant number of smaller clusters. At later times, however, when the exchange network is fully fragmented, $p(s)$ is well described by a power law decay with exponent approximately equal to $2$. This value effectively matches the critical exponent characterizing the distribution of cluster sizes in two-dimensional percolation ($\simeq 187/91=2.0549\ldots$ \cite{stauffer2018introduction}). 

\begin{figure*}
    \centering
    \includegraphics[width=\textwidth]{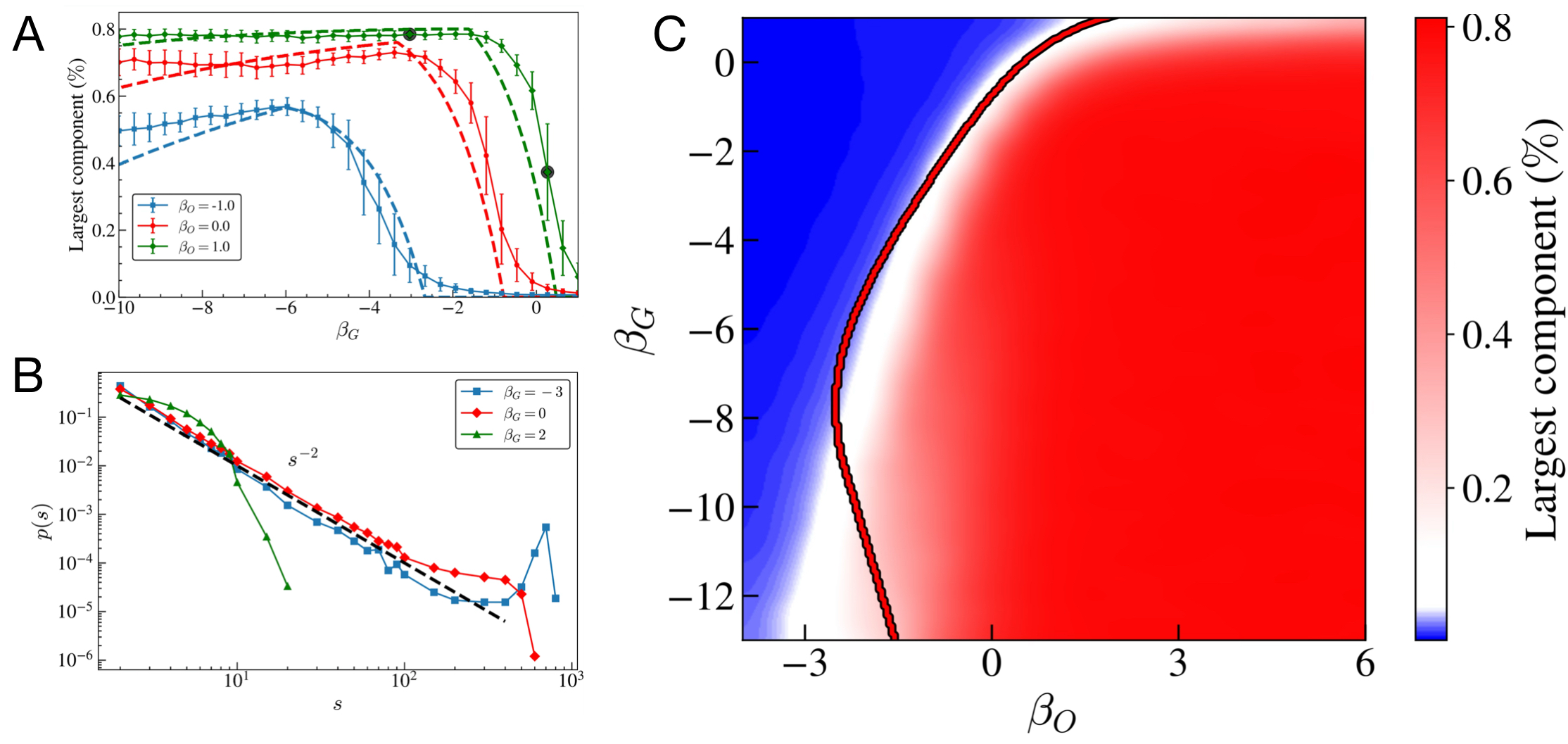}
\caption{{\bf Percolation cross-feeding transitions in the model. }(A) Average size of the largest connected component (in terms of the fraction of nodes belonging to it) as a function of $\beta_G$, for different values of $\beta_O$. Results from simulations of a two-dimensional system are represented by the markers with errors (continuous lines being a guide for the eye), while analytical results from the mean-field theory are shown as dashed lines. One sees that analytical results predict a sharp transition, as opposed to the crossover observed in simulations. 
(B) Distributions of cluster sizes for 3 different values of $\beta_G$ and $\beta_O=1$. The dashed black line displays the power-law behaviour
$p(s) \sim s^{-2}$. 
(C) Critical line for the existence of an extensive cluster in the $(\beta_O,\beta_G)$ plane as derived from mean-field theory. The background color map represents the relative size of the largest cluster from simulations of a two-dimensional system. Right of the critical line, the exchange network is system-spanning and dense; vice-versa, left of the line it is sparse and rich of small motifs. Again, the sharp transition predicted by mean-field theory turns into a smooth crossover in finite-dimensional simulations.}
    \label{fig3}
\end{figure*}

\subsection*{Mean-field theory: critical line for the emergence of an extensive  exchange cluster}


To rationalize these findings mathematically, let us focus on one cell and partition the space around it in circular stripes of width equal to the cell's radius $R$. In this way, the surface area of the $k$-th stripe equals $(2 k +1) \pi R^2$. If cells are distributed homogeneously in space with density $\rho$, the number of cells in the $k$-th stripe is a Poisson random variable with average given by $\mu_k=(2k+1) \pi R^2 \rho $. 
If we now let $x$ denote the probability that our cell belongs to the infinite percolating cluster (i.e. that it is connected to it by at least one other cell by a non-zero exchange flux) and assume that the lactate fluxes of individual cells are independent and identically distributed random variables, 
we have 
\begin{multline}\label{jhfddfghjk}
1-x = \\
\prod_{k\geq 1} \sum_{n\geq 0}  e^{-\mu_k}  \frac{\mu_k^n}{n!} \sum_{l=0}^{n} {n \choose l} (1-x)^l x^{n-l}  P_k^{n-l} ~~,
\end{multline}
where $P_k$ stands for the probability that the cell is not exchanging lactate with a cell in the $k$-stripe. 
To see why (\ref{jhfddfghjk}) holds (i.e. why the right-hand side is the complement of $x$), note first that the term  $(1-x)^l x^{n-l}P_k^{n-l}$ represents the probability that, if $n$ cells inhabit the $k$-th stripe from our cell, the latter is not connected to $l$ of them and is  exchanging lactate below the threshold with the remaining $n-l$ cells. By weighting each such factor with ${n\choose l}$ and summing over $l$ we are therefore computing the probability that the cell is effectively isolated from all of the $n$ cells in the $k$-th stripe. We next sum over $n$ by considering that it is a Poisson random variable. And finally impose that this must hold for all stripes, i.e. take the product over $k$. Now note that
\begin{equation}
\sum_{l=0}^{n} {n \choose l} (1-x)^l x^{n-l} P_k^{n-l} 
= (1-x+x P_k)^n ~~,
\end{equation}
and that
\begin{multline}
\sum_{n\geq 0}  e^{-\mu_k}  \frac{\mu_k^n}{n!} (1-x+x P_k)^n \\ = e^{-\mu_k} \sum_{n\geq 0} \frac{\mu_k^n(1-x+xP_k)^n}{n!}\\
=e^{-\mu_k x (1-P_k)} ~~,
\end{multline}
which leads to
\begin{equation}
1-x=\prod_{k\geq 1} e^{-\mu_k x (1-P_k)}~~.
\end{equation}
This equation is equivalent to the 
standard mean-field percolation formula \cite{stauffer2018introduction}
\begin{equation}
\sum_{k\geq 1} \mu_k (1-P_k) = -\frac{\log (1-x)}{x}~~,
\end{equation}
which implies the emergence of a giant cluster when the sum on the left-hand side exceeds 1. 

To proceed, we consider a continuum approximation for the quantity $\sum_{k\geq 1} \mu_k (1-P_k)$. Denoting by $Q(r)$ the probability that two cells at distance $r$ are connected (i.e. they exchange lactate with a flux above the threshold $\theta$) and by $\rho$ the density of cells, we have 
\begin{equation}\label{gfdhgjtkl}
\sum_{k\geq 1} \mu_k (1-P_k) \simeq \rho\int_{R}^L Q(r) \,2\pi r\,dr ~~.\end{equation}
We can now approximate $Q(r)$ as the probability that both the (incoming) flux of the absorbing cell ($u_{\text{abs}}$) and the (outgoing) flux of the emitting cell ($u_{\text{emi}}$) at the boundary of the absorbing cell exceed the significance threshold, the latter being damped by the factors $e^{-r/\xi}$ and $R/r$ that take into account dispersion (with a characteristic screening length scale $\xi$) and absorption, respectively. We furthermore assume that the distributions of lactate fluxes are Gaussian, with averages and variances given by the maximum entropy metabolic network model described in the previous section.  This implies that 
\begin{equation}
Q(r) \simeq 2\,\Phi(u_{\text{abs}}\leq -\theta)\,\Phi(u_{\text{emi}}\geq \theta e^{r/\xi} r/R)    ~~,
\end{equation}
where $\Phi$ denotes the cumulative Gaussian distribution function. The above expression can be inserted into (\ref{gfdhgjtkl}) to retrieve the critical line from the condition $\log(1-x)/x=-1$  as a function of $\beta_G$ and $\beta_O$, provided we have an expression for the screening length scale $\xi$. As a minimal assumption, we can posit that $\xi$ is a linear function of the average lactate flux, namely
\begin{equation}
\xi = A + B\langle u_L \rangle    ~~.
\end{equation}
The best fit of the simulation results (see Figure \ref{fig3}A below) yields for $A$ and $B$ the values $A\simeq 3.3 \cdot 10^{-2}$ and $B\simeq -1.3 \cdot 10^{-4}$.

Key results from mean-field percolation theory are showcased in Fig. \ref{fig3}. We first focus on the relative size of the largest component of the exchange networks, corresponding to $x$ in the mean-field theory (Fig. \ref{fig3}A). One clearly sees that the mean-field model qualitatively recapitulates results obtained in simulations for two-dimensional systems, albeit with a sharp dense-to-sparse transition instead of a smooth crossover. Likewise, the distribution of cluster size, obtained from simulations crossing the percolation line, displays qualitative agreement with results from simulations (Fig. \ref{fig3}B). Most notably, an algebraic decay with exponent close to 2 is quantitatively retrieved at intermediate values of $\beta_G$. Finally, Fig. \ref{fig3}C shows the critical line, defined by the condition $\log(1-x)/x=-1$, in the $(\beta_O,\beta_G)$ plane. Generally speaking, for any $\beta_O$, large enough values of $\beta_G$ drive the system into a sparse networked phase, while the global exchange network is dense otherwise. The mean-field critical line qualitatively recapitulates the behaviour of the relative size of the largest cluster as found in simulations, including a re-entrance that is predicted by the mean-field model and found in simulations at large negative $\beta_G$ (Fig. \ref{fig3}A). The transition observed in the exchange network is therefore well captured, at least qualitatively, by a simple mean-field percolation scenario.

\begin{figure*}[ht!!!!!!!!!!!]
    \centering
\includegraphics[width=.99\textwidth]{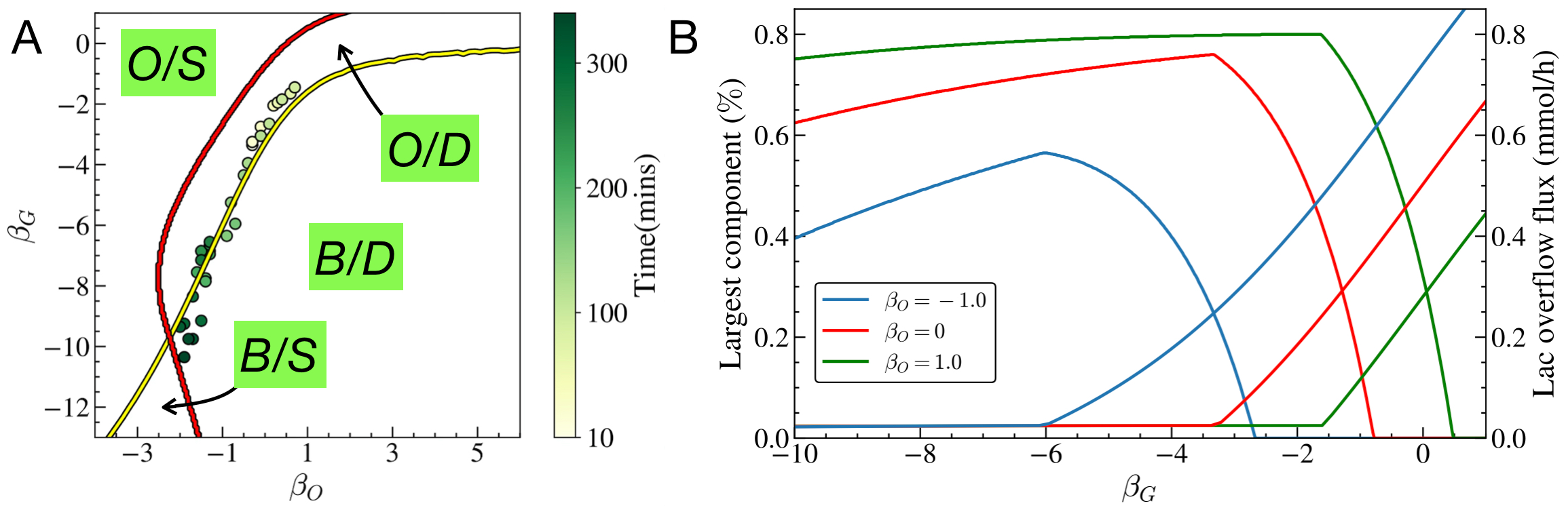}
\caption{{\bf Cross-feeding percolation and metabolic overflow.} (A): phase diagram of the multicellular metabolic network model in the plane $(\beta_O,\beta_G)$. Yellow line: overflow phase transition line dividing the plane into overflow (O) and balanced (B) phases, from \cite{narayanankutty2024emergent}. Red line: critical percolation line of lactate exchanges (from the mean-field approximation), dividing the plane into sparse (S) and dense (D) phases of the exchange web. Markers: values of $\beta_O$ and $\beta_G$ inferred from experimental data (color scale indicates the time step). The distinct regions correspond to overflow and sparse (O/S); overflow and dense (O/D); balanced and sparse (B/S); balanced and dense (B/D). (B) Overflow curves (average lactate flux, right-side axis) vs percolation curves (relative size of the giant component from mean-field theory, left-side axis) as a function of $\beta_G$ for three different values of $\beta_O$ (i.e. along three vertical lines in the phase diagram of panel (A)).}
    \label{fig4}
\end{figure*}

\subsection*{Comparison of cross-feeding percolation and metabolic overflow}

We can now re-visit overflow metabolism in the light of the percolation phase transition scenario derived above. In \cite{narayanankutty2024emergent}, again by combining mathematical modeling and statistical inference, it was shown that, for any $\beta_G$, the mean lactate output by cells (equivalent to the lactate spillover in the medium) behaves markedly differently at large $\beta_O$ compared to low $\beta_O$: in the former phase, no spillover in the medium (i.e. no acidification) occurs, while in the latter one is due to observe significant accumulation of lactate in the environment (Warburg effect). The line separating the two regimes in the $(\beta_G,\beta_O)$ plane is shown in Fig. \ref{fig4}A, along with the line for the existence of a dense cluster and with the values of $\beta_G$ and $\beta_O$ describing the empirical population (see \cite{narayanankutty2024emergent} for details). 
Fig. \ref{fig4}B displays instead the behaviour of the relative size of the largest component and the lactate overflow flux along 3 vertical lines with constant $\beta_O$ as a function of $\beta_G$. 

One first notes that the mean-field approximation suggests the existence of four regions, depending on whether the exchange network is dense or sparse and on whether it is balanced (i.e. the lactate spillover is small) or not. While an overflow-sparse regime could have been expected based on the fact that aberrant cells should independently optimize their metabolism for growth (and hence lactate excretion), the system appears to evolve towards balanced states in an overflow-dense regime that is harder to explain within cell-autonomous models.

In addition, one can see explicitly how the reduction of lactate spillover is accompanied by the sparsification of the exchange network. By staying close to the critical overflow line population are likely to optimize the trade-off between the environmental costs of pollution (high in the overflow phase) and the regulatory costs of coordinating metabolisms (high in the balanced phase). The present results now suggest that different structures for exchange networks are associated to optimal trade-offs in different phases. In particular, contrary to early adaptation phases where system-scale exchange networks dominate, cellular behaviour in balanced states appears to rely on multiple smaller communities of cells that keep microenvironmental acidification under control. 

Note also how, in Fig. \ref{fig4}B, the lactate overflow transition occurs when the size of the largest percolation cluster is maximal as a function of $\beta_G$.

Because, over time, the population tends towards states that maximize the energy yield on glucose (i.e. inferred values of $\beta_G$ and $\beta_O$) \cite{narayanankutty2024emergent}, we can hypothesize that adaptation involves two intertwined processes. On shorter timescales, cells focus on the trade-off between environmental and regulatory costs, leading to dense exchange networks; on longer timescales, however, they aim at becoming more energy-efficient individually, implying short-range coordination and reduced acid excretion \cite{palsson2015systems}. Both aspects are likely essential to formulate even a minimal dynamical model of how a population of cells adapts to an environment.

\section*{Conclusions} 

The main result of this work is that, in a population of cells adapting to an environment, cells exchange lactate through a diffusion-mediated network whose topology varies in time as the adaptation proceeds. Here we approached the problem of exchange network reconstruction quantitatively. In particular, 
\begin{enumerate}
\item[(i)] we devised a computationally  efficient network reconstruction technique combining Voronoi tessellation of the culture with simple equilibrium results for absorption probabilities;
\item[(ii)] using this method, we showed that the experimental population undergoes a two-stage dynamics: following inoculation, cells first build up a dense and intense exchange network comprising up to 70\% of them; subsequently, as the adaptation proceeds, the network dilutes, eventually leading to the formation of smaller disconnected clusters of interacting cells;
\item[(iii)] we demonstrated that the key topological features of the inferred exchange networks (from the size of the largest component to the distribution of cluster sizes) are quantitatively reproduced by a minimal (Maximum Entropy-like) model that accounts for the metabolic behaviour of cells and their interplay with the environment; the model is controlled by just two parameters, $\beta_G$ and $\beta_O$, fixing on average mean glucose and oxygen consumptions of cells;
\item[(iv)] by focusing on the general properties of the model, we showed that, as $\beta_G$ and $\beta_O$ change, exchange networks generically undergo a transition from a densely connected phase characterized by the presence of a large cluster of cells involved in mutual exchanges to a disconnected phase in which exchanges occur within small disconnected clusters. Such a transition can be studied quantitatively using methods of percolation theory, leading to a critical line in the $(\beta_O,\beta_G)$ plane separating a dense exchange regime from a sparse one. Real populations appear to evolve close to the crossover line as time progresses.
\end{enumerate}
To rationalize these findings, we resorted to a time-independent modeling framework in which cells are only required to meet maintenance thresholds, exchanges are driven by diffusion, and changes in the ensuing inter-cellular network are modulated by the average level of carbon and oxygen intake by cells, measured respectively by the parameters $\beta_G$ and $\beta_O$. We found that, for any given value of $\beta_G$, a dense network is expected for sufficiently large values of $\beta_O$. Conversely, the exchange network is sparse and dominated by small motifs for small enough $\beta_O$. Such a behaviour can be associated to a percolation-like phase transition in the mean-field limit (see the phase diagram of Fig. \ref{fig3}C). Simulation results on a finite-dimensional system however effectively agree with this scenario. Therefore, as a population modulates its mean carbon and oxygen intakes over time, the network gradually adapts. Indeed network inference results from empirical data show that the exchange matrix starts dense and becomes more and more sparse over time, approaching a state that resembles, at least statistically, a critical percolation network.

Interestingly the critical percolation threshold for the exchange network formation and the phase transition line of the metabolic overflow do not coincide. This situation is similar to that found in studies attempting to derive a geometric interpretation for statistical observables in standard spin systems or in gas-liquid phase transition \cite{kertesz1989existence,schioppa1998coniglio}. In a nutshell, an equivalence between certain thermal ensemble-averaged quantities and percolative quantities is known to exist, upon a specific re-definition of the percolation model \cite{coniglio1980clusters}. This observation has in turn inspired further theoretical insights \cite{fortunato2003critical} as well as the development of cluster algorithms for sampling near criticality \cite{swendsen1987nonuniversal}. While beyond the scope of the present work, an analogous re-definition of percolation in the context of interacting cell models could, in our view, be equally fruitful. 

To conclude, we remark that theoretical approaches to exchange network reconstruction problems from input/output fluxes have a long history in network theory \cite{squartini2017maximum,squartini2018reconstruction}. In the most involved instances (e.g. the reconstruction of  weighted exchange networks from single-node data), rather non-trivial statistical mechanics is involved \cite{mastrandrea2014enhanced}. In the present case, however, the fact that diffusion is the basic exchange mechanism offers a strong proxy for the exchange probability, thereby allowing to figure out simpler solution method(s). On the other hand, our work effectively addresses an instance of a multi-layer network \cite{bianconi2018multilayer}, in which a large-scale exchange network arises from the more or less coordinated activity of a large number of single-agent networks (the metabolic networks of individual cells). As the key role of cell-to-cell communication becomes more and more clear through experiments, the theoretical problem of linking them to intracellular networks increasingly acquires importance. The methods developed here can provide a solid starting point to this analysis whenever diffusion is the main driver of inter-cellular signaling.

\bibliographystyle{unsrt}
\bibliography{name}

\appendix

\section*{Appendix}

\subsection*{Monte Carlo algorithm}

After setting a step length $r_0>R$, for each emitting cell $i$ the following algorithm is employed:
\begin{enumerate}[leftmargin=*]
\item Initialize a random walk at position $\mathbf{r}_i$ where lies a cell $i$ having $u_i < 0$ (i.e. an emitting cell);
\item Pick a random angle $\alpha$ uniformly in $[0,2\pi]$ and move the diffusing molecule a distance $r_0$ in the direction given by $\alpha$. If the molecule reaches a boundary it is neglected and a new emission begins;
\item If the molecule arrives within $r_0$ of a cell $j$ having $u_j>0$ (i.e. of an absorbing cell), it is absorbed by $j$ with probability proportional to $u_i$ 
If molecules reach the boundary they are lost (we assume open boundary conditions).
\item Iterate steps 1-3 $M\gg 1$ times for each emitter $i$ (with $M$ the number of diffusing molecules) and compute $M(j|i)$, the number of molecules emitted at $i$ absorbed at $j$ for each $i$ and $j$.
\item The probability $P(j|i)$ is approximated by
\begin{equation}
P(j|i)\simeq\frac{M(j|i)}{\sum_i M(j|i)}~~,
\end{equation}
where $\sum_i M(j|i)\equiv M(j)$ represents the number of walkers absorbed at $j$.
\item Finally, the intercellular exchange between $i$ and $j$ will be given by 
\begin{equation}
    u_{i\to j}=u_j P(j|i)~~.
\end{equation} 
\end{enumerate}


\subsection*{Metabolic model parameters}

\begin{table}[h!]
\centering
\begin{tabular}{|c|c|c|} 
\hline
\textbf{Constant} & \textbf{Symbol} & \textbf{Value} \\ \hline
Maximum oxygen uptake & $U_O$ & $3$mmol/gh \\ \hline
Maximum glucose uptake & $U_G$ & $1$mmol/gh \\ \hline
ATP Maintanance flux & $M$ & $1$mmol/gh \\ \hline
Cell radius & $R$ & $10 \mu$m \\ \hline
Number of cells & $N$ & $10^3$ \\ \hline
Observed system length & $L$ & $1.3$mm \\ \hline
Critical lactate threshold & $\phi_c$ & $0.025$mmol/gh \\ \hline

\end{tabular}
\caption{Parameters used in the constraint-based metabolic model, see \cite{narayanankutty2024emergent} for further details.}
\label{tab:example}
\end{table}

\subsection*{Robustness of the network reconstruction}
\begin{figure}
    \centering
\includegraphics[width=.47\textwidth]{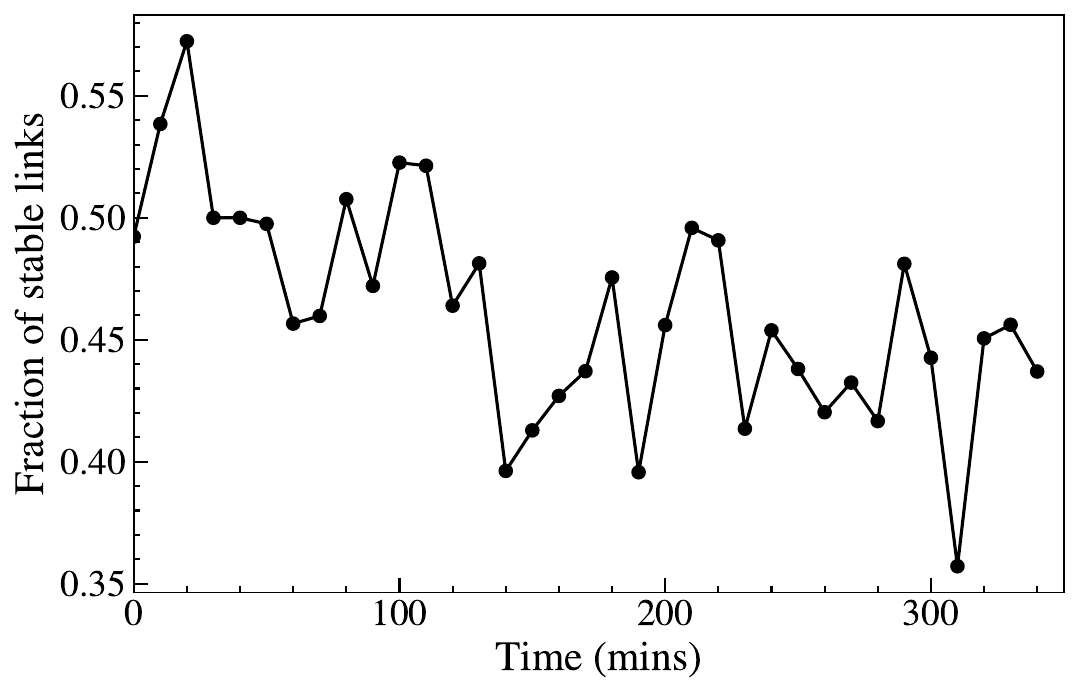}
\caption{Fraction of links that are stable in the network reconstruction as a function of time.}
    \label{figA}
\end{figure}
The estimation of single-cell metabolic fluxes in \cite{onesto2022probing} comes with errors that in turn affect the uncertainty of the reconstructed network, in particular limiting the ability to fully determine single links.
In particular, starting from the pairwise approximation described by (\ref{mfu}),   
one finds that 
\begin{multline}
\left(\frac{\delta u_{ij}}{u_{ij}}\right)^2 = \frac{1}{Z_j^2}  \sum_{k\ne i:u_k>0} \left( \frac{\delta u_k}{d_{kj}}\right)^2 \\
+ \left( 1 + \frac{u_j}{u_{ij}} \right)^2 \left(\frac{\delta u_i}{d_{ij}}\right)^2 +
\left(\frac{u_i}{u_{ij}}\right)^2\left( \frac{\delta u_j}{d_{ij}} \right)^2~~.
\label{eq:errprog}
\end{multline}
Fig. \ref{figA} shows the fraction of links that are stable in the  reconstruction, i.e. whose average value minus the propagated error overcome the threshold. One sees that roughly half of the individual links are robustly determined. In turn, collective statistical features like the size of the largest connected component display robust behavior across instances, as shown by the size of the error bars in Fig. \ref{fig2}D.

\end{document}